\documentclass[preprint2]{aastex}

\slugcomment{To appear in the Astronomical Journal, November 2003}

\shorttitle{Candidate damped Lyman-$\alpha$ absorbers}
\shortauthors{Lacy et al.}

\begin{document}


\title{Imaging and spectroscopy of galaxies associated with 
two $z\sim 0.7$ damped Lyman-$\alpha$ absorption systems}


\author{Mark Lacy \altaffilmark{1,2,3}, 
Robert H.\ Becker \altaffilmark{2,3},  
Lisa J.\ Storrie-Lombardi \altaffilmark{1}, 
Michael D.\ Gregg\altaffilmark{2,3}, 
Tanya Urrutia \altaffilmark{3}, Richard L.\ White\altaffilmark{4}} 

\altaffiltext{1}{SIRTF Science Center, MS 220-6, California Institute of 
Technology, 1200 E.\ California Boulevard, Pasadena, CA 91125; 
mlacy@ipac.caltech.edu, lisa@ipac.caltech.edu}

\altaffiltext{2}{IGPP, L-413, Lawrence Livermore National Laboratory, 
Livermore, CA 94550; bob@igpp.ucllnl.org,
gregg@igpp.ucllnl.org}

\altaffiltext{3}{Department of Physics, University 
of California, 1 Shields Avenue, Davis, CA 95616; 
urrutia@physics.ucdavis.edu }

\altaffiltext{4}{Space Telescope Science Institute, 3700 San Martin Drive,
Baltimore, MD 21218; rlw@stsci.edu }

\begin{abstract}
We have identified galaxies near two quasars which are at the redshift
of damped Lyman-$\alpha$ (DLA) systems in the UV spectra of the quasars. 
Both galaxies are actively forming stars. One 
galaxy has a luminosity close to the break in the local galaxy luminosity 
function, $L^{*}$, the other is significantly fainter than
$L^{*}$ and appears to be interacting with a nearby companion. Despite the 
strong selection effects favoring spectroscopic identification of the 
most luminous DLA galaxies, many of the spectroscopically-identified DLA
galaxies in the literature are sub-$L^{*}$, suggesting that the majority
of the DLA population is probably sub-$L^{*}$, in contrast to Mg{\sc ii}
absorbers at similar redshifts whose mean luminosity is close to $L^{*}$. 
\end{abstract}



\section{Introduction}

Damped Ly$\alpha$ (DLA) absorbers in quasar spectra are the class of 
absorbers with the highest column density of neutral hydrogen ($N_{HI} \geq 
2\times 10^{20} {\rm cm^{-2}}$). They are important as it can be shown 
that they contain the bulk of the neutral hydrogen content of the Universe 
(e.g.\ Storrie-Lombardi \& Wolfe 2000). 
The similarity of the column densities of DLAs to those through the disks
of spiral galaxies 
have lead some authors to consider them as the direct progenitors of 
present-day spirals (e.g.\ Prochaska \& Wolfe 1998). 
However, dwarf galaxies and low surface brightness
galaxies (LSBGs) also can have similar columns 
(Boissier, Peroux \& Pettini 2003), and metallicity arguments support
dwarf galaxies (Pettini et al.\ 1999).
Theorists have interpreted them in the context of hierarchical galaxy 
formation models as sub-units which will eventually merge to
form $L^{*}$ galaxies today (Haehnelt, Steinmetz \& Rauch 1998;
Maller et al.\ 2001) or as dwarf 
galaxies in which star formation is supressed by supernova feedback
(Efstathiou 2000). 

The reason for the proliferation of theories for the nature of DLAs is 
that reliably identifying the galaxies responsible for the absorption,
even at $z<1$, has proven to be very difficult. Imaging studies of $z<1$
absorbers, e.g.\ Le Brun et al.\ (1997); Rao et al.\ (2002) indicate
that a range of galaxy types
might be responsible for producing DLAs, but few candidate DLAs have measured
redshifts, so reliable constraints on the nature of the DLA population  
remain poor. Paradoxically, 
Mg{\sc ii} absorbers, with much lower column densities on average, have been
easier to identify.
They seem to be sampling the normal galaxy population, with a mean luminosity
of $\sim L^{*}$ at $<z> = 0.6$
(Steidel, Dickinson \& Persson 1994; Steidel et al.\ 2002). Possible reasons
for the difficulty of identifying the galaxies causing DLA 
systems include small impact 
parameters, low stellar luminosities or low surface brightnesses.
So far only six $z<1$ DLAs have been 
spectroscopically-confirmed (and all of these are at $z<0.6$). Several 
are dwarf or low surface brightness galaxies, the best-studied
example of which is SBS~1543+5921 (Bowen, Tripp \& Jenkins 2001), 
but the DLA absorber 
of AO~0827+24 is a fairly normal, luminous spiral (Steidel et al.\
2002). Many more 
suggested identifications have been made on the basis of proximity to the 
quasar and consistent photometry, but the reliability of these is unclear.

DLAs are easy to identify in the 
spectra of high redshift ($z\stackrel{>}{_{\sim}} 2$) quasars as they show up 
clearly in observed optical spectra, but they are hard to identify at lower 
redshift. By adopting the strategy of taking UV spectra with the 
{\em Hubble Space Telescope } ($HST$) of the highest 
rest-frame equivalent width ($EW_0$) Mg{\sc ii} absorbers in the literature, 
Rao \& Turnshek (2001) were able to identify a sample of lower 
redshift ($z\sim 0.3-1.7$) DLAs (see also Boisse \& Bergeron\ 1998). 
High $EW_0$ Mg{\sc ii} absorbers also typically show high $EW_0$ Fe{\sc ii}
absorption lines, and these have been hypothesised to be better tracers
of cool gas than Mg{\sc ii} (Bergeron \& Stasi\'{n}ska 1986).
We have therefore 
used the FIRST Bright Quasar Survey (FBQS; White et al.\ 2000; Becker et 
al.\ 2001) 
to select a similar sample with $EW_0$ (Fe{\sc ii} 2600\AA)$\geq 1$\AA. 
We have taken spectra of 26 of the sample with STIS on the {\em HST} 
to look for DLA systems. We have 
followed up this sample using the Keck telescope, taking near-infrared 
images in sub-arcsecond seeing, medium-resolution spectra of the quasars, 
and low-resolution spectra of the candidate absorbers. 
We are mid-way through spectroscopic follow-up of the FBQS sample, and 
have identified four galaxies with redshifts the same as that of the 
Fe{\sc ii} absorbers. Of these four, however, only two are associated with 
absorbers which are definitely damped. These two galaxies, near the 
quasars FBQS 0051+0041 (R.A.\ $00^{\rm h} 51^{\rm m} 30\fs 49$ 
Dec.\ $+00^{\circ} 41^{\rm m} 49\farcs91$ (J2000); absorber redshift, 
$z_{\rm abs}=0.74$; quasar redshift, $z_{\rm Q}=1.19$; apparent magnitude
$B=18.5$) and 
FBQS 1137+3907 (R.A.\ $11^{\rm h} 37^{\rm m} 09\fs 46$ 
Dec.\ $+39^{\circ} 07^{\rm m} 23\farcs 60$ (J2000);
$z_{\rm abs}=0.72$; $z_{\rm Q}=1.02$; $B=18.4$), 
form the subject of this paper. 

We assume a cosmology with 
$\Omega_{\rm M}=0.3, \Omega_{\Lambda}=0.7$ and $H_0=70 {\rm kms^{-1}Mpc^{-1}}$
throughout. Where we have referenced galaxy luminosities, $L$, to the 
present-day values of the break in the luminosity function, $L^{*}$ we 
have used $L^*$ values of Norberg et al.\ (2002) in $B$, 
Lin et al.\ (1996) in $R$ and Loveday (2000) in $K$, adjusted to our passbands
and cosmology as necessary ($M_B^{*}=-20.2$, $M_R^{*}=-21.1$, 
$M_K^{*}=-24.4$). The $k$-corrections have been derived using a
model 5Gyr old Sd galaxy from the Pegase library (Rocca-Volmerange \& 
Fioc 1997). We picked this model as representative of a moderately young,
star-forming galaxy, a decision motivated by the discovery of [O{\sc ii}] 
emission from the candidate DLA galaxies.

\section{Imaging Observations}

\begin{figure}
\plotone{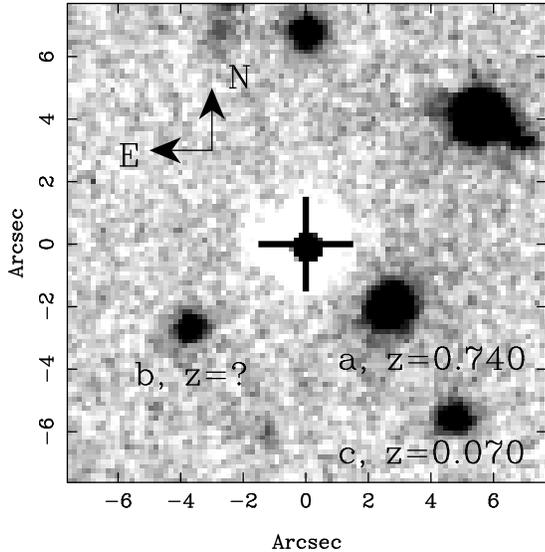}
\caption{The field of FBQS 0051+0041 in $I$. The quasar has been partially
subtracted from near the center of the image, and its position is marked
with a cross. The DLA candidate is galaxy `a'} 
\end{figure}

\begin{figure}
\plotone{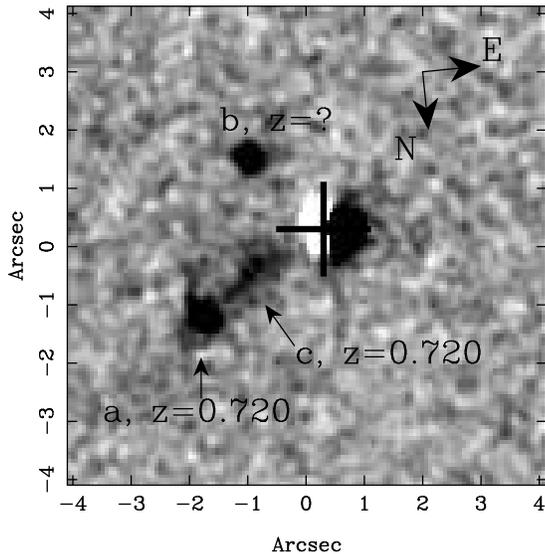}
\caption{The field of FBQS 1137+3907 in $K^{'}$. The quasar has been
partially subtracted from near the center of the image, and its position is marked
with a cross. Galaxies `a' and `c' are most probably associated with the
DLA system.} 
\end{figure}

FBQS~0051+0041 was imaged with the Echellette 
Spectrograph and Imager (ESI) on Keck-II and 
FBQS~1137+3907 with the Near-Infrared Camera 
(NIRC) on Keck-I. Details of the observations are given in
in Table 1. A sequence of short (3-5min) dithered exposures of each object
were taken. After dark subtraction and flat-fielding with dome flats the 
data were analysed using {\sc dimsum} and {\sc xdimsum} in IRAF, 
respectively. The {\sc dimsum} package performs two 
passes on the data. In the first pass, a background image is made 
from a median of the dithered frames and the data are combined with integer
pixel shifts to make masks which are used to mask out objects when 
constructing an improved 
sky background image for the second pass. In the second pass, the  
improved background is subtracted
from the images before they are shifted and coadded, and a magnification
factor may be applied to the data. The pixel scale of the ESI 
image was sufficiently small compared to the seeing that no magnification was
used in the second pass for FBQS~0051+0041.
The operation of the {\sc dimsum} package is explained in more detail 
in Stanford, Eisenhardt \& Dickinson (1995). The {\sc xdimsum} package
is a modification of {\sc dimsum} by L.\ Davis at NOAO. Amongst other 
improvements it runs significantly faster and allows non-integer 
magnification factors in the second pass. We used a factor of 2.3 for the 
FBQS~1137+3907 image.

Photometry for FBQS~0051+0041 was obtained using observations of the 
Landolt (1992) standard 107-215, and that for FBQS~1137+3907 from observations
of the standard star P266-C from Persson et al.\ (1998). Both quasar fields
were imaged in photometric conditions. The $5\sigma$ limiting magnitudes
in 2-arcsec diameter apertures are $I=25.4$ for the FBQS~0051+0041 field and
$K^{'}=21.9$ for the FBQS~1137+3907 field.

We performed approximate 
subtractions of the quasars using point spread functions derived from
either from a star in the same image (FBQS~0051+0041) or from other quasars
observed on the same night (FBQS~1137+3907). The PSF was fitted with a 
least-squares fit using code developed for fitting quasar host galaxies
(Lacy et al.\ 2002). The particular algorithm employed here subtracted the
PSF such that the mean value within a radius $r_{\rm in}$ from the center
of the quasar was set to zero. The values of $r_{\rm in}$ used were 
0\farcs 6 for FBQS~0051+0041 and 0\farcs 3 for FBQS~1137+3907. Neither PSF 
subtraction was very satisfactory, as we made no serious
attempt to obtain good, unsaturated PSFs for the quasars. They 
have successfully removed the
wings of the PSFs, however. The resulting images are shown 
in Figures 1 and 2.

\section{HST Spectroscopy}

The quasars were observed with the STIS with the MAMA detectors and the G230L
grating as part of program 9051. The 0\farcs 2 slit was used, giving a 
spectral resolving power of 330 at the redshifted wavelength of 
Ly$\alpha$, corresponding to a velocity resolution of 
$\approx 900 {\rm kms^{-1}}$.
Each quasar was observed for two 1100s exposures, 
the second exposure being dithered $2^{''}$ along the slit. The UV spectrum
of FBQS~1137+3907 shows it to be a high-ionization 
broad-absorption line (BAL) quasar, with a BAL trough blueshifted by 
$13000{\rm kms^{-1}}$.  
Fortunately, the Lyman-$\alpha$ absorption line is clear of the BAL troughs, 
falling just to the red of the O{\sc vi} trough in the quasar
spectrum. The spectra are shown in Figures 3 and 4.

%%% LISA  
A detailed discussion of the HST spectra and measurements of the damped
absorbers are provided in a companion paper (Storrie-Lombardi, et.~al. 2003, 
in preparation). The column densities of the damped absorbers were
initially estimated from measurements of the equivalent widths of 
the absorption lines with the redshift determined from the low ionization
metal lines from our ground-based spectra. Profiles were then fit iteratively 
to determine the best fit column density and redshift for HI absorption
feature. The column densities should be accurate to $\pm 0.1$ and the 
redshifts to $\pm 0.005$.  Though higher resolution spectra covering multiple
lines in the Lyman series would be ideal for this purpose,  
the substantial HST time that would be involved in obtaining them makes 
such observations 
impractical for large samples of objects. At higher redshifts, we 
have shown via observation and
simulation that moderate resolution spectra provide an excellent estimate 
for the measured HI column densities (e.g. Storrie-Lombardi et.~al. 1996; 
Storrie-Lombardi \& Wolfe 2000),
which gives us high confidence that the features in these spectra 
are damped and are not blends of lower column density lines. 
%%% LISA  

\begin{figure}
\plotone{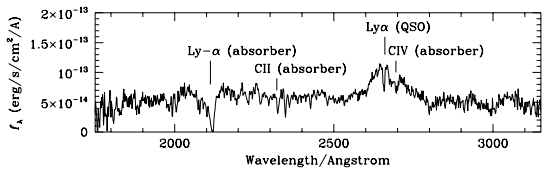}
\caption{STIS spectrum of FBQS 0051+0041. The quasar ($z_{\rm Q}=1.19$)
and absorber ($z_{\rm abs}=0.74$) features
are labelled.}
\end{figure}

\begin{figure}

\plotone{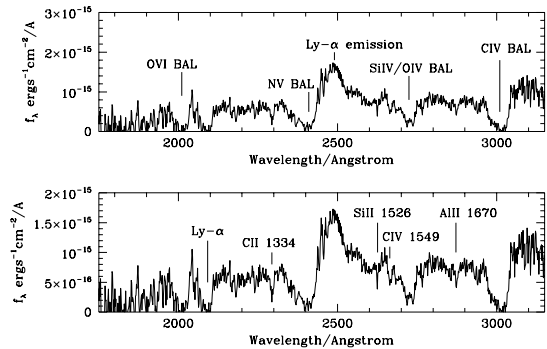}

\caption{The STIS spectrum of FBQS 1137+3907. The upper plot has the 
quasar Ly$\alpha$ emission (at $z_{\rm Q}=1.02$) and BAL features labelled, 
the lower plot has absorption features associated with the DLA system 
($z_{\rm abs}=0.72$) labelled.}
\end{figure}

\begin{figure}
\plotone{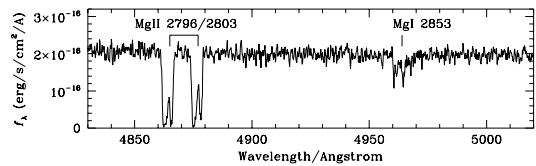}
\caption{The Keck ESI spectrum of FBQS 0051+0041 in the region of the 
Mg{\sc ii} absorption lines.}
\end{figure}

\begin{figure}

\plotone{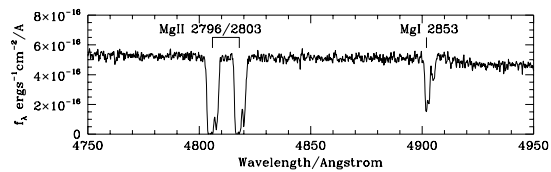}

\caption{The Keck ESI spectrum of FBQS 1137+3907 in the region of the 
Mg{\sc ii} absorption lines.}
\end{figure}

\section{Keck spectroscopy}

The candidate absorbers were observed with the Low Resolution
Imaging Spectrometer (LRIS) on Keck-I using the 
300 line mm$^{-1}$, 5000\AA$\;$blaze grating in the red arm
and a 1-arcsecond slit (see Table 2). This combination results in a spectral
resolving power of $\approx 500$, corresponding to a velocity resolution
of $\approx 600 {\rm kms^{-1}}$. The wavelength range covered was from 
$\approx 5000-10000$\AA. 
 
In addition, observations of the quasars were made with ESI on Keck-II to 
obtain medium-resolution spectra of the absorption lines. The Fe{\sc ii}
2586 and 2600\AA$\;$lines fell in a noisy region near an order overlap
in the ESI spectra, but the
Mg{\sc ii} 2796 and 2803\AA$\;$lines were detected at 
high signal-to-noise in both quasars. FBQS 0051+0041 was observed with ESI 
on 2000 December 31 with a $0\farcs 75$ slit, and FBQS 1137+3907
on 2001 March 20 with a $1^{''}$ slit, yielding resolving powers of 
5400 and 4100 respectively. The ESI spectra of the quasars
in the region around Mg{\sc ii} are shown in Figures 5 and 6, and
the LRIS spectra of the galaxies at the absorber redshifts 
in Figures 7 and 8.

\section{The absorber of FBQS 0051+0041}

The {\em HST} spectrum shows that this absorber is a borderline DLA, with a 
column density just above the division between Lyman-limit and DLA systems
(Table 3). Similarly, the $EW_0$ of the Fe{\sc ii} absorption only just 
satisfies our selection criteria. 

The candidate absorber galaxy, galaxy `a' in Figure 1,
seems, at first sight, to be a fairly normal, star forming
galaxy, offset by 24kpc from the quasar (Figures 1 and 7). 
The spectrum, however, shows a small mis-match
between the [O{\sc ii}]3727 redshift of 0.740 and the absorption line redshift
from the Ca II H+K lines (0.738), 
though the signal-to-noise ratio of the absorption
lines is low. Two distinct velocity components are also 
seen in the low-ionization Mg{\sc ii}, Mg{\sc i} and Fe{\sc ii} quasar 
absorption lines in our ESI spectrum, one at $z=0.7393$ and
one at $z=0.7402$ (i.e.\ a $160{\rm kms^{-1}}$ velocity
difference). Whether these two components correspond to two galaxies,
the candidate and a dwarf companion nearer the quasar, or reflect internal
gas motions in the candidate absorber is unclear.

\section{The absorber of FBQS 1137+3907}

This is a much higher column density absorber than that of FBQS~0051+0041. 
The {\em HST} spectrum shows several strong UV 
metal absorption lines in addition to the damped Ly$\alpha$ line (Figure 4).

The LRIS slit was placed to cover both candidate absorbing 
galaxies, $a$ and $c$, and line
emission is seen from both of them (barely resolved from each other) 
at the wavelength corresponding to
[O{\sc ii}] 3727 at $z=0.7185 \pm 0.0003$ (Figure 9). There is also a marginal
detection of H$\beta$ at the same redshift (Figure 8).
Although we see no other features from $a$ or 
$c$, it seems unlikely that the strong emission line can be anything
other than [O{\sc ii}] at the absorber redshift, given its high
equivalent width and the 
lack of lines nearby in the spectrum. We also took a spectrum with the slit 
aligned with object `b'. This showed that `b' had blue optical continuum 
emission, but no strong features from which to obtain a redshift. As it is 
further from the quasar than `c' and has no features at wavelengths
corresponding to the absorption redshift we assume that it is unrelated to 
the DLA system.

\onecolumn

\begin{table*}
\caption{Imaging of the FBQS DLA fields}
\begin{tabular}{lccccc}
\tableline
\tableline
Quasar field & Telescope/Instrument & Filter  &Observation & Total integration &
Seeing \\
             &                      &        & date (UT)   & time (minutes)    &
(arcsec)        \\
\tableline
FBQS~0051+0041    & Keck-II/ESI &$I$& 2001 June 22      &20&0.8\\
FBQS~1137+3907    & Keck-I/NIRC &$K^{'}$& 2002 May 25   &27&0.4\\
\tableline
\end{tabular}
\end{table*}

\begin{table*}
\caption{Observed properties and spectroscopic observations of the FBQS 
DLA galaxies}

{\scriptsize
\begin{tabular}{lcccccccccc}
\tableline
\tableline
Quasar field & $z_{\rm gal}$&Mag. &Impact    &Impact     & Abs.&$\Delta M_{k}$&$L/L^{*}$&UT Date& Slit PA &Exposure \\
             &              &     &parameter &parameter  & mag.&&         & (deg) &  &time (s) \\
             &              &     &(arcsec)  & (kpc)     &     &&         &       &  & \\
\tableline			 
FBQS~0051+0041a    & 0.740   & $I=22.0$   &3.3 &24  & $M_{\rm B}=-20.0$&-0.11&0.8&2001 Sept 25&26 &$2\times 1200$\\
FBQS~0051+0041b    &   ?     & $I=23.3$   &4.6 &?   &  -     &  -      &&&  -& - \\
FBQS~0051+0041c    & 0.070   & $I=23.2$   &7.2 &9.6 & $M_{\rm R}=-13.9$&-0.15&0.001&2001 Sept 25&26 &$2\times 1200$\\
FBQS~1137+3907a    & 0.7185  &$K^{'}=19.5$&2.5 &18  & $M_{\rm K}=-23.0$&-0.08&0.3&2002 Dec 31 &126&$2\times 1800$\\
FBQS~1137+3907b    & ?       &$K^{'}=20.4$&1.8 &?   & -                &-&&2002 Dec 31 &228&$2\times 1800$\\
FBQS~1137+3907c    & 0.7185  &$K^{'}=19.8$&1.5 &11  & $M_{\rm K}=-22.8$&-0.08&0.2&2002 Dec 31 &126&$2\times 1800$\\
\tableline
\end{tabular}
}
Notes: $\Delta M_{k}$ is the amount that was added to the absolute 
magnitudes to account for the difference in the flux density of the galaxy
between the observed band and the band the absolute magnitude was calculated 
in. This correction depends
only on the assumed galaxy SED and reduces to the usual $k$-correction when the two bands are the same.
\end{table*}

\begin{table*}
\caption{Details of the DLA systems}
\begin{tabular}{llccccc}
\tableline			 
\tableline			 
Quasar & $z_{abs}$& MgII $EW_0$ &FeII $EW_0$&  lg($N_{\rm HI}$ \\
       &          & (2796) &    (2600)  &  ${\rm (cm^{2})}$)\\
%       &           &       &            &                  \\
\tableline				      
FBQS~0051+0041    & 0.740  &	2.4   & 1.0  	& 20.4 $\pm 0.1$ \\
FBQS~1137+3907    & 0.719  &	3.0   & 2.5 	& 21.1 $\pm 0.1$ \\
\tableline
\end{tabular}
%%% LISA -- added errors to column density

\end{table*}

\begin{table*}
\caption{Physical properties of galaxies with spectroscopic redshifts 
consistent with $z_{\rm abs}$}

{\scriptsize
\begin{tabular}{llccccccll}
\tableline				      
\tableline				      
Absorber             & $z_{\rm gal}$&${\rm lg(N_{HI}}$&Impact   &Observed&Absolute  &$\Delta M_k$&$L/L^{*}$&Type&Refs.\ \\
candidate            &              &${\rm (cm^{2})}$)&parameter&mag.     &mag.  &            &         &    &       \\
                     &              &&(kpc)    &         &      &            &         &    &       \\
\tableline			 
SBS 1543+5921        & 0.009&20.4& 0.4     &$R=16.3$          &$M_{\rm R}=-16.7$& -   & 0.02& LSBG     &9,10 \\
OI 363 (B0738+313) G1& 0.221&20.9&      20 &$R=20.8$          &$M_{\rm R}=-19.4$&0.2  & 0.2 &dE    &3,4 \\
FBQS 1137+3907c      & 0.720&21.1&      11 & $K^{'}=19.8$     &$M_{\rm K}=-22.8$&-0.1 & 0.2 &spiral?  &1 \\
PKS 1127-145         & 0.312&21.7&    $<$7 &$R=21.6^{c}$      &$M_{\rm B}=-19.3$&-0.2 & 0.4 &LSBG     &7,8 \\
PKS 1243-072A        & 0.437&$\approx 20.5^{b}$&12&$I=21.3$   &$M_{\rm R}=-20.1$&0.2  & 0.4 &starburst/Seyfert 2&8 \\
FBQS 0051+0041a      & 0.740&20.4      &24 &$I=22.0$          &$M_{\rm B}=-20.0$&-0.11& 0.8&spiral?   &1\\
B2 0827+243 G1       & 0.526&20.3&      36 &$I=20.3$          &$M_{\rm R}=-21.6$&0.2  & 1.6 &spiral   &5,6 \\
AO 0235+164 A$^{a}$  & 0.524&21.7$^{b}$&13 &$m_{\rm 785}>20.3$&$M_{\rm R}>-22.6$&0.3&$<4$&Seyfert 1&2 \\
\tableline
\end{tabular}
}
% col dens are 0235: 5.0e-21 (Chengalur & Kanakar MN 2000), OI363 7.9e20 (RT)
% 0827 2.0e20 (RT) PKS1127 51e20(RT) PKS1243 ~5e20 (7) and HS1543 20.35 (8)

\noindent
Notes: $^{a}$ object A1 in ref.\ 2 is closer to the quasar but has no 
spectrum, if it 
is at the absorber redshift it has $(L/L^{*})_{B}=1.0$, A has a bright nuclear
component, and the luminosity of the underlying galaxy is uncertain; we have
quoted upper limits based on the sum of the host and AGN fluxes. $^{b}$ column
density from 21cm line absorption, dependent on the assumed spin temperature. $^{c}$ sum
of components 1-4 of Rao et al.\ (2003). See Table 3 for the definition of 
$\Delta M_k$.
References: (1) this paper; (2) Burbidge et al.\ 1996; (3) Turnshek et al.\
2001; (4) Cohen 2001; (5) Steidel et al.\ 2002; (6) Rao et al.\ 2003; 
(7) Lane et al.\ 1998; (8) Kanekar et al.\ 2002; (9) Reimers \& Hagen 1998; (10) Bowen et al.\ 2001

\end{table*}

\twocolumn
 
%Faint line emission, 
%extended about $1^{''}$ from the quasar is also seen 
%corresponding to [O{\sc ii}] 3727 at  $z=1.02$, 
%the redshift of the quasar (there is 
%no emission line at the quasar redshift which would correspond to 
%[O{\sc ii}] 3727 at $z\approx 0.72$). 
%There is thus the possibility that $a$ or,
%more probably, $c$, are associated with the quasar host galaxy, however, 
%there is no strong emission line in $K^{'}$ that would lead to an extension 
%of the quasar image, and the PA of the radio source (along which such 
%emission might be expected to align, e.g.\ Ridgway \& Stockton 1997) 
%is 38 deg., about 90 deg.\ from the PA joining $a$ to the quasar. 

The ESI spectrum shows a complicated velocity profile, with two to four 
distinct low-ionization absorption components seen, depending on which line
is examined. The two deepest features, visible in all the lines, are at 
$z=0.7184$ and $z=0.7193$ (i.e.\ a $160{\rm kms^{-1}}$ velocity
difference). Thus the [O{\sc ii}] emission is most likely to be 
associated with the deeper, less redshifted absorption component.

\begin{figure}
\plotone{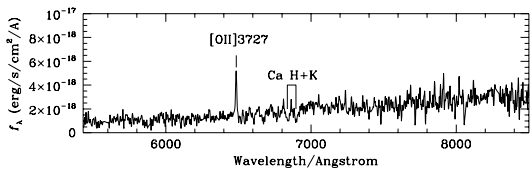}
\caption{The spectrum of galaxy `a' ($z=0.740$) near FBQS 0051+0041.}
\end{figure}

\begin{figure}

\plotone{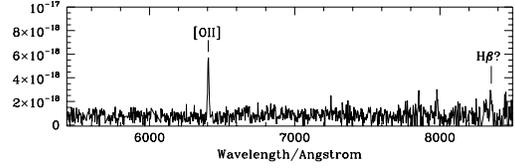}

\caption{The sum of the extracted 
spectra of galaxies `a' and `c' near FBQS 1137+3907.}
\end{figure}

\begin{figure}

\plotone{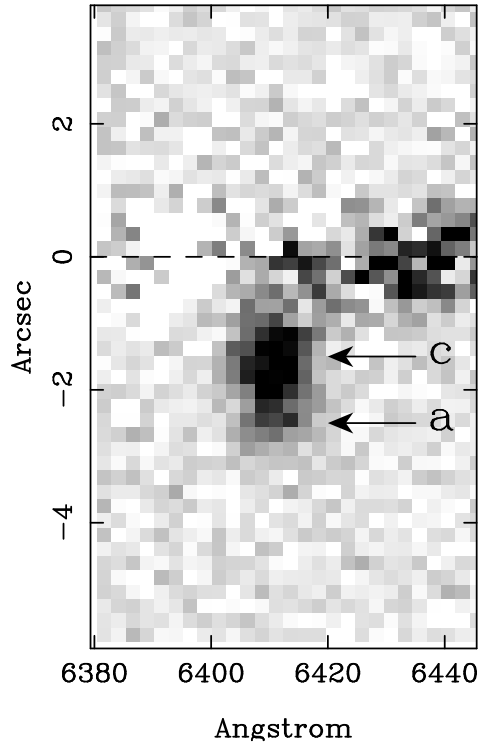}

\caption{The 2D spectrum of galaxies `a' and `c' near FBQS 1137+3907
in the region of the [O{\sc ii}] emission line in the galaxies. The 
trace of the quasar was averaged for 30 pixels either side of the region
shown and the mean subtracted from this portion of the image. The center
of the unsubtracted quasar trace is indicated by the dashed line at 
0 arcsec. The seeing FWHM was 0\farcs 92 (4.3 pixels), so `a' and `c' are 
only marginally resolved.}
\end{figure}

\section{Discussion}

%The importance of obtaining spectroscopic information on the absorber 
%candidates is illustrated by the preliminary results of our FBQS study...
%Reliable photometric redshifts are hard to obtain due to the proximity of the
%bright quasar.

The DLA absorber candidate close to FBQS 0051+0041 has a luminosity of 
approximately $L^{*}$, whereas that associated with FBQS 1137+3907 is about 
$0.2 L^{*}$. The galaxy types are unclear. Both are apparently forming stars,
possessing both blue continua and [O{\sc ii}] emission lines, and the 
complicated velocity structures of the Mg{\sc ii} absorption lines seen in the
ESI spectra, might be indicative of starburst-induced
winds (e.g.\ Ellison et al.\ 2003). Further evidence for this is the
offset between the emission and absorption 
redshifts in FBQS 0051+0041a, if real. Thus there is some evidence that both 
may be starburst galaxies. However, the continuum slope of FBQS 0051+0041a
is a little too red unless the starburst is reddened by dust, and  
FBQS 1137+3907c has a high axial ratio, suggesting a spiral 
rather than an irregular galaxy (though the shape of `c' may 
also be due to it being composed of a chain of knots). 
Higher spatial resolution multicolor 
imaging will be required to determine the nature of these galaxies.

Table 4 summarizes the properties of the eight known
candidate DLA galaxies at $z<1$ 
which spectroscopy has shown to lie at the absorber redshift. 
Selection effects will mean that the absorber population for which 
spectroscopic identification is possible will be biased. Observational 
requirements for successful spectroscopy, namely an impact 
parameter $\stackrel{>}{_{\sim}}1$ arcsec and an apparent magnitude brighter
than $R\sim 24$, will naturally favor massive galaxies. For our 
FBQS sample discussed in the introduction, and from which FBQS 0051+0041
and FBQS 1137+3907 were drawn, we have so far 
only successfully identified candidates and obtained 
spectra with redshifts consistent with those of the absorber system 
for four out of thirteen of our high $EW_0$ absorption systems. 
In addition, there is another subtle selection 
effect from the way most (though not all) of the DLAs 
in Table 4 have been selected, namely on the 
basis of high $EW_0$ metal absorption lines.
This will tend to bias the sample towards higher metallicity, 
more massive, galaxies. Working against these two effects is  
the possibility that many of the more massive DLA galaxies are dusty
enough to significantly extinguish the light from the background quasar, and
cause the quasar to drop out of optically-selected samples. This may tend to 
produce a bias towards relatively dust-free, less massive hosts, and  
may explain why the Mg{\sc ii} absorbers, with their generally 
higher impact parameters and lower H{\sc i} column densities, 
are observed to have luminosities around $L^*$. However,
the results of the survey of radio-loud quasars of Ellison et al.\ (2001) 
suggest that this is not an important effect. Any such effect will also
be much reduced for FBQS quasars as they are selected on the Palomar 
$E$-plates with a relatively red ($O-E \sim B-R <2.0$) color selection 
criterion. Thus, despite most of the selection effects favouring 
successful spectroscopic identification of the most luminous absorbing DLA 
galaxies, it is clear from Table 4 that many are 
sub-$L^{*}$, and this suggests that the majority of the DLA population 
is probably sub-$L^{*}$.

Both our DLA candidates appear to be actively forming stars, consistent 
with them being relatively gas-rich galaxies. A large fraction of
the known spectroscopically-confirmed DLA systems at low redshift listed in 
Table 4 appear to be merging or interacting in some way, for example, 
FBQS~1137+3907c, PKS~1127-145 and PKS~1243-072~A
(though obviously the statistics are poor at present, and few apparent
pairs have been spectroscopically confirmed). 
This is intriguing in the light of theories in which DLA
galaxies are dwarf galaxies which merge to become massive galaxies today 
(Haehnelt et al.\ 1998; 
Maller et al.\ 2001). Though these theories are more applicable to the 
high redshift systems, the lower redshift DLAs may represent the 
tail end of this activity.

\acknowledgments

We thank an anonymous referee for helpful comments.
This work was partly carried out at the Jet Propulsion Laboratory, 
California Institute of Technology, under contract with the National 
Aeronautics and Space Administration (NASA), and 
partly under the auspices of the U.S.\ Department of Energy, National
Nuclear Security Administration by the University of California, 
LLNL, under contract No.\ W-7405-Eng-48, 
with additional support from NASA grant HST-GO-09051.01-A and 
NSF grant AST-00-98355 (University of California, Davis). 
Some of the data presented herein were obtained at the W.M.\ Keck
Observatory, which is operated as a scientific partnership among the 
California Institute of Technology, the University of California and NASA.
The observatory was made possible by the generous financial support of the
W.M.\ Keck Foundation. The authors wish to acknowledge the 
significant cultural
role and reverence that the summit of Mauna Kea has always had within the 
indigenous Hawaiian community. This research has also 
made use of the NASA/IPAC Extragalactic Database (NED) which is operated by 
the Jet Propulsion 
Laboratory, California Institute of Technology, under contract with NASA.

%% Generally speaking, only the figure captions, and not the figures
%% themselves, are included in electronic manuscript submissions.
%% Use \figcaption to format your figure captions. They should begin on a
%% new page.

%% No more than seven \figcaption commands are allowed per page,
%% so if you have more than seven captions, insert a \clearpage

%% There must be a \figcaption command for each legend. Key the text of the
%% legend and the optional \label in curly braces. If you wish, you may
%% include the name of the corresponding figure file in square brackets.
%% The label is for identification purposes only. It will not insert the
%% figures themselves into the document.
%% If you want to include your art in the paper, use \plotone.
%% Refer to the on-line documentation for details.

%% Tables should be submitted one per page, so put a \clearpage before

%% Two options are available to the author for producing tables:  the
%% deluxetable environment provided by the AASTeX package or the LaTeX
%% table environment.  Use of deluxetable is preferred.
%%

%% Three table samples follow, two marked up in the deluxetable environment,
%% one marked up as a LaTeX table.

%% In this first example, note that the \footnotesize command has been
%% used to shrink the table so it will fit on one page. Note also that
%% the \label command needs to be placed inside the \tablecaption.


\begin{thebibliography}{}

\bibitem[]{} Becker, R.H.\ et al.\ 2001, ApJS, 135, 227
\bibitem[]{} Bergeron, J.\ \& Stasi\'{n}ska, G.\ 1986, A\&A, 169, 1
\bibitem[]{} Boisse, P., Le Brun, V., Bergeron, J.\ \& Deharveng, J.-M.\ 1998, A\&A, 333, 841
\bibitem[]{} Boissier, S., Peroux, C.\ \& Pettini, M.\ 2003, MNRAS, 338, 131
\bibitem[]{} Bowen, D.V., Tripp, T.M.\ \& Jenkins, E.B.\ 2001, ApJ, 121, 1456
\bibitem[]{} Burbidge, E.M., Beaver, E.A., Cohen R.D., Junkkarinen, V.T.\
\& Lyons, R.W.\ 1996, AJ, 112, 2533
\bibitem[]{} Cohen, J.G.\ 2001, AJ, 121, 1275
\bibitem[]{} Efstathiou, G.\ 2000, MNRAS, 317, 697
\bibitem[]{} Ellison, S.L., Yan, L., Hook, I.M., Pettini, M., Wall, J.V.\
\& Shaver, P.\ 2001, A\&A, 379, 393
\bibitem[]{} Ellison, S.L., Mall\'{e}n-Ornelas, G.\ \& Sawicki, M.\ 2003,
ApJ, in press (astro-ph/0302147)
\bibitem[]{} Haehnelt, M.G., Steinmetz, M. \& Rauch, M. 1998, ApJ, 495, 647
\bibitem[]{} Kanekar, N., Athreya, R.M.\ \& Chengalur, J.N., 2002, A\&A, 382, 
838
\bibitem[]{} Kulkarni, V., et al.\ 2001, ApJ, 551, 37
\bibitem[]{} Landolt, A.U.\ 1992, AJ, 104, 340
\bibitem[]{} Lane, W., Smette, A., Briggs, F., Rao, S., Turnshek, D. \& 
Meylan, G., 1998, AJ, 116, 26
\bibitem[]{} Le Brun, V., Bergeron, J., Boiss\'{e}, P.\ \& Deharveng, J.M.\ 1997, A\&A, 321, 733
\bibitem[]{} Lin, H., Kirshner, R.P., Shectman, S.A, Landy, S.D., Oemler, A.,
Tucker, D.L.\ \& Schechter, P.L.\ 1996, ApJ, 464, 60 
\bibitem[]{} Loveday, J.\ 2000, MNRAS, 312, 557
\bibitem[]{} Maller A.H., Prochaska, J.X., Somerville, R.S.\ \& Primak, J.R.\ 2001, MNRAS, 326, 1475
\bibitem[]{} Norberg, P., et al.\ 2002, MNRAS, 336, 907
\bibitem[]{} Persson, S.E., Murphey, D.C., Krezeminski, W., Roth, M.\ \& 
Rieke, M.J.\ 1998, AJ, 116, 2475
\bibitem[]{} Pettini, M., Ellison, S.L., Steidel, C.C. \& Bowen, D.V. 1999, ApJ, 510, 576 
\bibitem[]{} Prochaska, J.X.\ \& Wolfe, A.M.\ 1998, ApJ, 507, 113
\bibitem[]{} Rao, S.\ \& Turnshek, D.A.\ 2000, ApJS, 130, 1
\bibitem[]{} Rao, S., Nestor, D.B., Turnshek, D.A., Lane, W.M., Monier, E.M.\
\& Bergeron, J.\ ApJ, submitted (astro-ph/0211297)
\bibitem[]{} Ridgway, S.E.\ \& Stockton, A.N.\ 1997, AJ, 114, 511
\bibitem[]{} Rocca-Volmerange, B.\ \& Fioc, M.\ 1997, A\&A, 326, 950
\bibitem[]{} Reimers, D.\ \& Hagen, H.-J.\ 1998, A\&A, 329, L25
\bibitem[]{} Stanford, S.A., Eisenhardt, P.R.M.\ \& Dickinson, M.\ 1995, ApJ,
450, 512
\bibitem[]{} Steidel, C., Dickinson, M.\  \& Persson, E.\ 1994, ApJ, 437, L75
\bibitem[]{} Steidel, C., Kollmeier, J.A., Shapley, A., Churchill, C.W., 
Dickinson, M.\ \& Pettini, M.\ 2002, ApJ, 570, 526
\bibitem[]{} Songaila A., Cowie L.L., Hu E.M., Gardner J.P., 1994, ApJS, 94, 461
\bibitem[]{} Storrie-Lombardi, L.J, McMahon, R.G., Irwin, M.J. \& Hazard, C. 1996, ApJ, 468, 121
\bibitem[]{} Storrie-Lombardi, L.J. \& Wolfe, A.M. 2000, ApJ, 543, 552
\bibitem[]{} Turnshek D.A., Rao, S., Nestor, D., Lane, W., Monier, E., 
Bergeron, J.\ \& Smette, A.\ 2001, ApJ, 553, 288
\bibitem[]{} Warren, S.J., Moller, P., Fall, S.M.\ \& Jakobsen, P.\ 2001, MNRAS, 326, 759
\bibitem[]{} White, R.L.\ et al.\ 2000, ApJS, 126, 133
\bibitem[]{} Zwaan M., Briggs, F.H.\ \& Verheijen, M.\ 2002, in Extragalactic Gas at Low Redshift, eds J.S.\ Mulchaey \& J.\ Stocke, ASP Vol.\ 254, p.\ 169



\end{thebibliography}
\end{document}